# Time Symmetric Quantum Theory Without Retrocausality

Tim Maudlin, Department of Philosophy, New York University, New York, NY 10003

**Abstract**: In the recent paper "Is a Time Symmetric Interpretation of Quantum Theory Possible Without Retrocausality?", Mattew Leifer and Matthew Pusey argue that the answer to their title question is "no". Unfortunately, the central proof offered in the paper contains a fatal error, and the conclusion cannot be established. Interpretations of quantum theory without retrocausality can be time symmetric not only in the traditional sense but in Leifer and Pusey's supposedly stricter sense. There appear to be no prospects for proving any analogous theorem.

Section 1: The Set-Up

In a recent paper [1], Matthew Leifer and Matthew Pusey attempt to show that under certain conditions a time-reversible physical theory must incorporate a form of retrocausality. Such a result meets with the immediate objection that many time-reversible theories—including Newtonian mechanics, Maxwellian electrodynamics, Bohmian mechanics and Many Worlds theory—are time reversible but include no such retrocausality. The laws of these theories generate solutions successively from past initial conditions forward in time with no need (and indeed no place) for any causal input from the future. Leifer and Pusey respond that their notion of time-reversibility is not the usual one, and none of these commonly accepted examples of time-reversibility meet their stricter requirement. Having explicated their criteria for both time-reversibility and retrocausality, they mount an argument that the former (with plausible additional conditions) entails the latter.

The details of Leifer and Pusey's argument are challenging to follow. In the course of the argument they impose unusual conditions for counting a variable as an "input" or "output" of a process, and further restrict their attention to the sector of the theory in which there can be no signaling to either the future or the past. The purpose of this note is to show that the argument is invalid, and to trace down the source of the error in the derivation.

Quite a lot of the structure of the argument can best be understood by back-formation: the structural model that Leifer and

Pusey attempt to replicate is Bell's theorem. Many of the equations in this paper look just like equations that Bell used, and the fundamental mathematical result is Bell's. Leifer and Pusey characterize their result as a timelike version of Bell's result. While it is essential for Bell's argument that the two measurements involved occur at spacelike separation, in this argument the two supposedly analogous parts of the experiment (we cannot call them two measurements since one is a preparation and the other a measurement) occur at timelike separation. Leifer and Pusey try to motivate precisely the same structural constraints for a theory with no retrocausality in this timelike setting as Bell demonstrates for a local theory in a spacelike setting.

Already alarm bells ought to be sounding. Bell argues that in a local theory certain correlations between spacelike separated experiments cannot occur. That is because in a local theory (in Bell's sense) there can be no direct influence between spacelike separated events. But there is no corresponding general prohibition for influences between timelike separated experimental conditions: they can display whatever correlations you like. The input and output of the first experimental condition can be conveyed to the second one by usual subluminal means, and the second can then adjust itself by means of that information to yield the desired correlations.

Let's quickly recall what Bell did.[1] He considers a source or preparation that produces some complete physical state $\lambda$ with probability $\rho(\lambda)$. $\lambda$ then propagates forward in time and two space-like separated measurements $M_1$ and $M_2$ are made. Each measuring device can be set in N different ways, with the specification of the setting represented by the variables $x$ and $y$ respectively. The outcome of the $M_1$ measurement is $a$ and the outcome of $M_2$ $b$. $x$ and $y$ are treated as free variables since they can be chosen in whatever way one likes.

If there were no constraints on the physical theory, all we could say is that it should provide probabilities for the various possible measurement outcomes conditional on the specification of all the other physical conditions. For example, there should be a probability $p(a,b|x,y,\lambda)$ for getting the results $a$ and $b$ given that the complete physical state of the prepared state is $\lambda$ and the measurement settings are $x$ and $y$. The overall chance of getting the results $a$ and $b$ given settings $x$ and $y$ would then be $\int p(a,b|x,y,\lambda)\rho(\lambda)d\lambda$. Bell then argues that in a *local* theory $p(a,b|x,y,\lambda) = p(a|x, \lambda) \, p(b|y, \lambda)$. One step of the argument is that the free choice of *y* cannot be correlated with the outcome *a* in a local theory, as that would constitute a superluminal influence, and by parity of reasoning *b* cannot be correlated with *x*. So $p(a|x,y,\lambda) = p(a|x,\lambda)$ and $p(b|x,y,\lambda) = p(b|y,\lambda)$. The second step is argue that conditionalizing on $\lambda$ should screen off the outcomes *a* and *b* from each other since $\lambda$ contains a complete specification of everything in the common past of the two measurement events and hence, in a local theory, a specification of all common causes. In such a case $p(a|b,x,y, \lambda) = p(a|x,y, \lambda)$ and $p(b|a,x,y, \lambda) = p(b|x,y, \lambda)$. Since $p(a, b|x, y, \lambda) = p(a|b, x, y, \lambda) \, p(b| x, y, \lambda)$ we derive $p(a, b|x, y, \lambda) = p(a|x,\lambda) \, p(b|y,\lambda)$ and $\int p(a, b|x, y, \lambda) \rho(\lambda) d\lambda \,) \;=\; \int p(a|x, \lambda) p(b|y, \lambda) \rho(\lambda) d\lambda$. From this the inequality follows.

---

[1] For more detail, see [2].

The guts of Leifer and Pusey's argument contain a causal graph, i.e. a directed graph whose edges represent possible causal influences. In a theory without retrocausality the graph must be acyclic: if causes always precede their effects and effects never precede their causes and the causal influence relation is transitive, then the graph can contain no closed causal cycles. Leifer and Pusey employ a very general model of an experiment, which they call the *PTM* model. *P* stands for *preparation*, *T* for *transformation* and *M* for *measurement*. As related in conventional time order, the experimenter prepares a state, which then may be transformed in some way, and finally is measured. This allows Leifer and Pusey to represent such an experiment as follows:

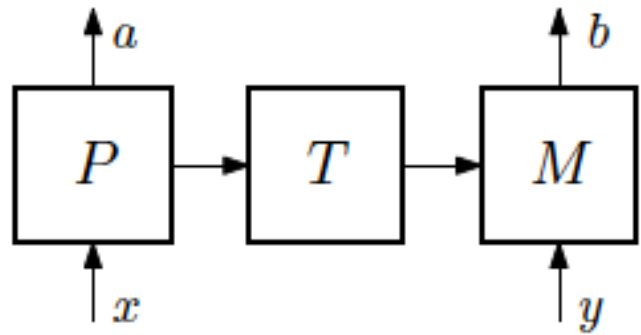

FIG. 1. Illustration of prepare-transform-measure experiment.

Figure 1: Influence Diagram of a Leifer and Pusey Experiment

Time in the diagram generally runs left-to-right. $x$ represents the experimenter's choice among possible preparation procedures. $a$ represents some visible output of the preparation. Since preparation procedures are not generally regarded as having any output save the prepared system, the exact purpose of $a$ is a bit obscure. The preparation procedure is followed by the transformation $T$, and then by the measurement $M$. The causal structure of the measurement is familiar: the experimenter freely chooses what to measure by adjusting $y$, and the measurement apparatus returns the result $b$. This structure is used twice by Bell, since two of the essential parts of his experiement are measurements.

Enlightenment about the purpose and significance of the preparation output $a$ is not forthcoming. A few pages after the above diagram we find the following passage:

> For example, in quantum theory a preparation $P$ is associated with a Hilbert space $\mathcal{H}_A$, a set of (unnormalized) density operators $\{\rho_{aA|x}\}$ on $\mathcal{H}_A$—one for each choice of $x$ and $a$—such that the ensemble average density operators $\rho_{A|x} = \Sigma_a\, \rho_{aA|x}$ are normalized $\mathrm{Tr}(\rho_{A|x}) = 1$. The preparation procedure starts with the experimenter choosing $x$. The preparation device then generates a classical variable $a$ with probability distribution $p(a|x) = \mathrm{Tr}(\rho_{aA|x})$, outputs $a$, and prepares the system in the corresponding (normalized) state $\rho_{aA|x}/p(a|x)$, which is subsequently fed into the transformation device. (p. 6)

Note that the role of $a$ in this story seems both unnecessary and redundant. Often, a preparation device has no variable corresponding to input from the experimenter at all. If we were doing Bell-type experiments, for example, we might build a device that always outputs a pair of electrons in the singlet state. Or if we were doing a 2-slit experiment we might want a source that reliably produces electrons with a fixed momentum. In these cases there would be no "classical variable" $a$ for the device to generate and output, and in addition no place for the experimenter to make any choice $x$.

Why, then, have *a* and the associated conditional probabilities for the preparation at all? And why include the box for the transformation *T*?: this part of the experiment could be assimilated into either the preparation procedure or the measurement at will. With *T* interposed between *P* and *M*, though, the figure reproduced above looks suspiciously like a schematic diagram of a Bell's inequality test with an arrow reversed. The structure of the Bell experiment is:

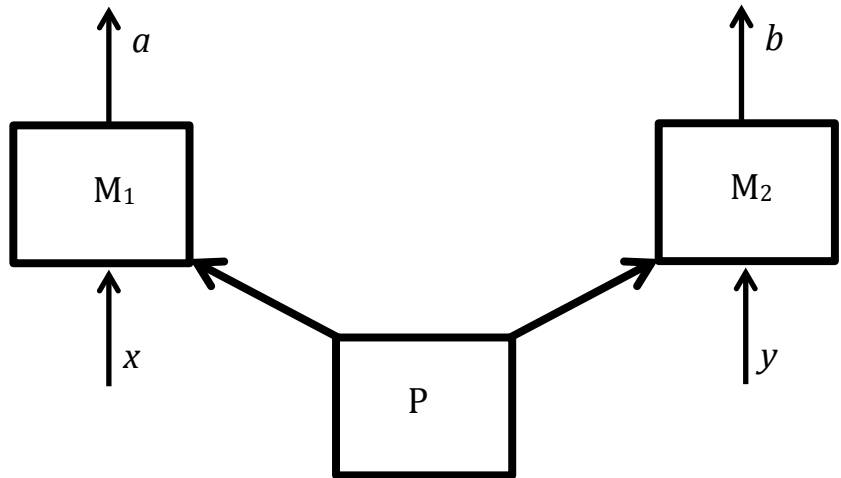

Figure 2: Influence Diagram of a Bell Test

The similarity of Figure 1 and Figure 2 is manifest, although time runs from left to right in Figure 1 and from bottom to top in Figure 2. The transformation phases are omitted in the Bell set-up. In this diagram *x* and *y* represent the choices of the experimenters about what angle of spin or polarization to measure and *a* and *b* represent the outcomes of the measurements. In Bell's case there is no freely adjustable input to the preparation *P*: the experimenter makes no change to *P* in the course of the experiment, nor does the preparation output anything save the system experimented on.

The key difference between the two diagrams is the reversal of one of the influence arrows. In Bell's case, no sequence of arrows runs from $M_1$ to $M_2$: neither measurement operation can influence the other. But in Figure 1 we have the chain of arrows running from *P* to *M*: the preparation can (of course) influence the result. Since Bell's restriction on local theories flows exactly from the lack of causal influence from one side to the other, it ought to be immediately counterintuitive to claim that anything like Bell's result can follow from a Figure 1 experimental structure.

Nonetheless, Leifer and Pusey desire to reproduce Bell's algebra. This is seen at a glance in the formulas, which look like they have come directly from Bell's paper. So we have, on the one hand, a basic causal structure that implies that there could not exist any analogous result to Bell's, and on the other hand the seeming reproduction of the probabilistic inferences that lead to his result. Something has to give. What is it?

Section 2: The Counterexample

Leifer and Pusey are trying to construct an argument parallel to Bell's. What Bell showed is that from two premises—Bell Locality and "Free Choice" (the scare quotes are to warn against the idea that this premise has anything at all to do with the notion of free will)—there follows a limitation on the sorts of correlations that can reliably occur for experiments done at spacelike separation. Leifer and Pusey's analog theorem is supposed to show that five premises—**No Retrocausality**, **λ-Mediation**, **Time Symmetry**, **Realism**, and **Free Choice**—imply that Bell's inequality cannot be violated by preparations and measurements done at timelike separation. They then show that Bell's inequality is indeed violated for some relevant quantum-mechanical preparations and measurements. It would then follow that at least one of the five premises must be rejected. Leifer and Pusey argue that **No Retrocausality** must go, leaving us with a world with backwards causation.

The claimed theorem, however, is mistaken. We will prove this by constructing a counter-example. With that in hand, we can track down where the error occurred.

The simplest experiment demonstrating a violation of Bell's inequality for measurements made at spacelike involves a choice between two measurements on each wing. For example, on the right side one makes a spin measurement at either 0° or 30°, and on the left either 0° or -30°. If the particles were prepared in the singlet state, then quantum theory predicts the following outcomes. When 0° is chosen on both sides, the outcomes always disagree: one is up and the other is down. When the angles differ by 30% (i.e. when the measurements are set at 0° and 30° or else at -30° and 0°), the particles give different outcomes 75% of the time. And when the measurements are set at -30° and 30°, the outcomes disagree 25% of the time. These statistical outcomes violate Bell's inequality.

Let's set up a physical theory that can produce these same statistics for experiments performed at timelike separation, with one the earlier experimental condition regarded as a preparation and the later as a measurement of the prepared state. Since the experiments are now at timelike separation, influence arrows can flow from the preparation to the measurement and there is no barrier for information to be conveyed from one experiment to the other. This makes reproducing the desired correlation child's play: the setting and outcome of the "preparation" is conveyed to the "measurement" which then reacts in the appropriate way to produce the desired correlations.

Here, then, is the model in complete detail. The preparation device has two settings for $x$, labeled 0 and 30, under the control of the experimenter. When the device is set one way or the other, it produces an outcome $a$ which is either "up" or "down". This outcome is produced at random with a probability of .5. (This step prevents the possibility of using the set-up to send signals.) The preparation device then emits a particle in a state $\lambda$ that contains information about both the experimental setting and the outcome. The measurement device receives $\lambda$ and also the setting $y$ on the measurement device. $y$ is either 0 or -30. Finally, given $x$, $a$, and $y$ the measurement device produces outcome $b$ ("up" or "down") with the right probabilities to yield the desired statistics.

Here is the model in more detail. There are 4 possible states of $\lambda$: <0,up>, <0,down>, <30,up> and <30,down>. After the preparation device has been set, it produces an up or down outcome with .5 probability and sends out the corresponding state of $\lambda$. The measurement device combines $\lambda$ with its own input of either 0 or -30 and then produces its output by the following scheme for $p(b|y,\lambda)$:

$p(\text{up}|0,<0,\text{up}>) = 0$
$p(\text{up}|0,<0,\text{down}>) = 1$
$p(\text{up}|0,<30,\text{up}>) = .25$
$p(\text{up}|0,<30,\text{down}>) = .75$
$p(\text{up}|-30,<0,\text{up}>) = .25$
$p(\text{up}|-30,<0,\text{down}>) = .75$
$p(\text{up}|-30,<30,\text{up}>) = .75$
$p(\text{up}|-30,<30,\text{down}>) = .25$

The probability that the measurement returns the outcome "down" given $y$ and $\lambda$ is just 1 less the probability that it returns "up".

We now have a theory that will violate the Bell inequalities. All we have to do is check that it satisfies all 5 of the conditions mentioned in the alleged theorem.

The theory satisfies **Free Choice** because the experimenters can choose the values of $x$ and $y$ however they like: there need be no correlations between $x$ and $y$. The theory satisfies **Realism** because on every run of the experiment the system is described by a definite state $\lambda$ that takes its value from a set of states $\Lambda$. In addition, on every run $a$, $b$, $x$, and $y$ take definite values. The theory satisfies **$\lambda$-mediation** because $\lambda$ mediates any correlations between the preparation and the measurement, i.e. $p(b| \lambda, a, x, y) = p(b| \lambda, y)$. This is trivial since $\lambda$ just is <$a$, $x$>. The theory satisfies **No Retrocausality** because there obviously is no retrocausality: everything happens in a normal causal sequence, and in particular the non-input variables $b$, and $\lambda$ are both conditionally independent of every input variable in their future (i.e. $y$) once they have been conditioned on all the variables in their past. Finally, the theory satisfies **Time Symmetry**. This is the most complicated condition, so we will go through it carefully.

Leifer and Pusey begin by defining what it is for a theory to be *operationally time symmetric*. All that means is that for every experiment *PMT* with inputs $x$ to the preparation and $y$ to the measurement and respective outputs $a$ and $b$ there exists an experiment *P'M'T'* with inputs $y$ to the preparation and $x$ to the measurement and outputs $b$ and $a$ respectively such that

$P_{P'M'T'}(b,a|y,x) = P_{PMT}(a,b|x,y)$.

*PMT* and *P'M'T'* are then called operational time reverses. In our case the symmetry of the correlational structure between the two sides guarantees that the theory is operationally time symmetric. Clearly we could have used the "preparation device" as a "measurement device" and the "measurement device" as a "preparation device" without altering the statistics for the joint outcomes. Leifer and Pusey prove (Theorem IV.3) that the no-signaling sector of a theory must be operationally time-symmetric. As noted above, the fact that the preparation uses a random selection of the output with .5 probability means that the system cannot be used for signaling, and so is operationally time-symmetric by their criterion.

With the definition of operational time symmetry in hand we can then define the stricter notion of *Ontological* Time Symmetry. Since this is the most critical property for our analysis, we will quote the definition in its entirety.

> In an ontic extension of an operational theory, the ontic extension ($P,M,T, \lambda$) of an operational theory ($P,M,T$) has an *ontological time reverse* if there exists another experiment ($P',M',T'$) with ontic extension ($P',M',T', \lambda'$), where $P'$ has the same set of inputs and outputs as $M$, $M'$ has the same set of inputs and outputs as $P$, and there exists a one-to-one map $f: \Lambda \to \Lambda'$, and
> 
> $$P_{P'M'T'}(b,a,f(\lambda)|y,x) = P_{PMT}(a,b,\lambda|x,y).$$
> 
> An ontic extension of an operational theory is *ontologically time symmetric* if every experiment has an ontological time reverse. (p. 12)

Our ontological extension of the operational theory $P_{PMT}(a,b|x,y)$ is easily seen to be ontologically time symmetric. Every experiment can be matched by an experiment in which the inputs of the preparation device have been relabeled -30 and 0. $f$ maps <0,up> and <0,down> to themselves and <30,up> and <30,down> to <-30,up> and <-30,down> respectively. The same substitutions in the probability function will clearly yield an ontological time reverse of the original experiment.

So our little theory is ontologically time symmetric according to Leifer and Pusey's definition. It satisfies all five of their conditions, and still can produce violations of Bell's inequality for correlations between the preparation and the measurement. The announced theorem is not a theorem.

Section 3: Diagnosis

What goes wrong with the argument?

Immediately after the definition of an ontologically time symmetric theory, Leifer and Pusey make the following rather curious remark:

> **Remark VI.2.** It is also true, mathematically at least, that if the ontic extension of an experiment has an ontological time reverse, then there also exists an extension in which $f$ is the identity. We can construct such an extension from an arbitrary one by identifying the ontic state spaces of the time reverse pairs of experiments and setting $p_{P'M'T'}(b, a, \lambda|x, y)$ in the new extension to the value of $p_{P'M'T'}(b, a, f(\lambda)|y, x)$ in the old extension. Physically, these two extensions may tell very different stories about what happens between preparation and measurement, but mathematically we can always assume that $f$ is trivial without loss of generality, and will do so in what follows.

It is not very easy to follow either the exact intent or the motivation of this remark. As written, we are told to construct a new model of the ontological time reverse that uses the same outputs as the original, switches the inputs, and replaces $f(\lambda)$ with $\lambda$. As mentioned above, in our example, $f(<0, up>) = <0, up>$, $f(<0, down>) = <0, down>$, $f(<30, up>) = <-30, up>$, and $f(<30, down>) = <-30, down>$. So to render $f$ the identity, we should systematically swap "30" in the primed model with "-30" and vice versa. But the main question is why in the world are we being asked to do this? Is it really so onerous to write $f(\lambda)$, or even $\lambda'$, instead of using $\lambda$ in both models? Without at least a prime on the $\lambda$ one is liable to lose track of which model is under discussion, and to mix up features that $\lambda$ has in one model with features it has in the other. Not surprisingly, it is this very confusion that lies at the heart of the error in the paper.

The error appears here:

**Lemma VIII.2**. *Let (P, M, T) be an experiment that has an operational time reverse. If its ontic extension satisfies **No Retrocausality** and **Time Symmetry** then*

$$p(\lambda|x, y) = p(\lambda), \qquad (16)$$
$$p(b|\lambda, x, y) = p(b|\lambda, y), \qquad (17)$$
$$p(a|\lambda, x, y) = p(a|\lambda, x). \qquad (18)$$

*Proof.* By **No Retrocausality**, the probabilities decompose as

$$p(a, b, \lambda|x, y) = p(b|\lambda, x, a, y)\, p(\lambda/a, x)\, p(a|x). \quad (19)$$

Using Bayes' rule, we have

$$p(\lambda|a, x) = \frac{p(a|\lambda, x) p(\lambda|x)}{p(a|x)}. \qquad (20)$$

Substituting this back into eq. (19) gives

$$p(a, b, \lambda|x, y) = p(b|\lambda, x, a, y)\, p(a|\lambda, x)\, p(\lambda|x). \quad (21)$$

Summing over a and b then gives $p(\lambda|x, y) = p(\lambda|x)$. By **Time Symmetry**, we also have the decomposition

$$p(a, b, \lambda|x, y) = p(a|\lambda, x, y, b)\, p(\lambda|b, y)\, p(b|y), \quad (22)$$

and applying the same argument to this gives $p(\lambda|x, y) = p(\lambda|y)$. We therefore have $p(\lambda|x, y) = p(\lambda|x) = p(\lambda|y)$, but this means that it cannot depend on either $x$ or $y$, so $p(\lambda|x, y) = p(\lambda)$. (p. 13)

Conclusion (16), that $p(\lambda|x, y) = p(\lambda)$, is manifestly wrong. The idea that there is some *unconditional* probability of $\lambda$ makes no sense. The whole point of $\lambda$ is to convey information about the preparation device from the preparation device to the measuring device. If $\lambda$ really were statistically independent of $x$ and $y$ then it could not play any important role in the experiment at all.

The mistake is easy to see. The authors derive (correctly) by **No Retrocausality** that in the *original* model by the value of $\lambda$ must be statistically independent of $y$ since the choice of $y$ lies to the future of the production of $\lambda$. And had they retained the nomenclature $f(\lambda)$, they would have accurately also remarked that in the ontological time reverse of the original model $f(\lambda)$ must be statistically independent of $x$ for the same reason. But due to their decision to make $f$ the identity they instead write that $\lambda$ is statistically independent of $x$. And from the premises that $\lambda$ is statistically independent of $y$ *in the original model*, and that $\lambda$ is statistically independent of $x$ *in the ontological time reverse model*, they draw the fallacious conclusion that $\lambda$ is statistically independent of *both x and y in both models*. Hence there must be some $p(\lambda)$: a probability of $\lambda$ *conditional on nothing* in both models. But that, of course, is nonsense. And it is hardly surprising that from such a premise one can derive that Bell's inequality cannot be violated. If $\lambda$ is conditionally independent of both $x$ and $a$ in the original model then it can convey no information at all about the preparation to the measurement device: the probability of getting any result from the measurement is independent of the preparation. In other words, the preparation plays no role in producing the outcome of the measurement. Such a model could not recover any correlations at all between the preparation and measurement, much less violations of Bell's inequality.

Needless to say, since the purported theorem is not a theorem the remaining discussion and conclusions of the paper are not reliable.

[1] M. Leifer and M. Pusey, *Proc. R. Soc. A* **473**: 20160607. http://dx.doi.org/10.1098/rspa.2016.0607

[2] T. Maudlin, *J. Phys. A: Math. Theor.* 47 (2014) 424010